\documentclass[conference]{IEEEtran}
\usepackage{amsmath,amssymb,amsthm,mathtools}
\usepackage{cite}
\usepackage{graphicx}
\usepackage{stfloats}
\usepackage{booktabs}
\usepackage{xcolor}
\usepackage[normalem]{ulem}
\definecolor{revisionblue}{RGB}{0,65,125}

\definecolor{revisionmaroon}{RGB}{112,24,48}
\definecolor{revisiongreen}{RGB}{0,85,45}
\definecolor{revisionpurple}{RGB}{76,25,112}

\usepackage{hyperref}
\hypersetup{hidelinks}
\newtheorem{theorem}{Theorem}
\newtheorem{proposition}{Proposition}
\newtheorem{lemma}{Lemma}
\newtheorem{remark}{Remark}

\newcommand{\E}{\mathbb{E}}
\newcommand{\Z}{\mathbb{Z}}
\newcommand{\norm}[1]{\lVert #1\rVert}
\newcommand{\Energy}{\mathcal{E}}

\begin{document}
\raggedbottom
\renewcommand{\topfraction}{0.9}
\renewcommand{\dbltopfraction}{0.9}
\renewcommand{\textfraction}{0.05}
\setlength{\dblfloatsep}{6pt plus 2pt minus 2pt}

\title{
Metropolis Random Walks Escape Deep Basins Faster in Higher Dimensions}

\author{
\IEEEauthorblockN{Hansini Ramachandran and Bhaskar Krishnamachari}
\IEEEauthorblockA{Department of Electrical and Computer Engineering\\
University of Southern California\\
Los Angeles, California\\
\{hramacha, bkrishna\}@usc.edu}
}

\maketitle

\begin{abstract}
 Simulated Annealing (SA) is a widely used stochastic optimization algorithm. It is of great interest to understand how its ability to escape local minima at a fixed temperature depends on geometry and dimension.
In prior work, we derived closed-form expressions for the expected basin escape time
of a constant-temperature SA process on a one-dimensional piecewise-linear landscape.
In this paper, we extend the analysis to $D$-dimensional cubic basins
defined by the infinity-norm energy $V(x)=\norm{x}_\infty$.
The walk proposes nearest-neighbor moves uniformly and escapes upon
first reaching the outer shell, that is, any site with
$\norm{x}_\infty=N$.
We construct a one-dimensional birth-death shell-quotient chain by
collapsing each concentric shell in the Metropolis electrical network
to a single node while retaining its total stationary mass.
We derive a closed-form expression for the mean escape time $Q_D(N)$ of this shell-quotient chain as a function
of the acceptance probability~$p$, the basin depth~$N$, and the dimension~$D$.
 For fixed $D$ and fixed $0<p<1$, the quotient escape time scales as
$\Theta(\alpha^N/N^{D-1})$ as $N\to\infty$, where $\alpha=1/p$.
Relative to $D=1$, this gives an asymptotic speedup of order $N^{D-1}$
within this family of cubic basins.
We prove that the true Metropolis mean escape time $H_D(N)$, without shell averaging, has
the same $\Theta(\alpha^N/N^{D-1})$ scaling, establishing a lower bound
via Rayleigh monotonicity and a matching upper bound through an explicit
low-energy flow construction using Thomson's principle.
 Numerical solutions that exploit the walk's symmetries quantify the
discrepancy between the quotient and true-chain means for $D=1,2,3$.
\end{abstract}

\section{Introduction}

Simulated Annealing (SA), introduced by Kirkpatrick, Gelatt, and
Vecchi~\cite{kirkpatrick1983}, is a widely used stochastic optimization
algorithm with applications ranging from Very Large Scale Integrated
(VLSI) circuit placement~\cite{sechen1985} to quantum circuit
compilation~\cite{molavi2025}.
It mimics the physical process of bringing a system to its minimum energy
state by controlling its temperature, using the Metropolis acceptance
rule~\cite{metropolis1953}.
 Prior works have applied convergence theory, spectral methods, and potential-theoretic
metastability to studying similar escape dynamics,
including in multidimensional settings~\cite{bovier2002,bovier2015}.
Here we seek an explicit discrete calculation that isolates the effect
of dimension on the time to escape a basin. 

In recent work~\cite{ramachandran2026}, we developed a finite-time
analytical framework for constant-temperature SA by studying a
piecewise-linear energy landscape in one dimension.
We modeled SA as a discrete-state Markov chain and derived closed-form
expressions for the expected time to escape a single linear basin,
both in the single-basin and two-basin settings. 
However, that analysis was restricted to one-dimensional landscapes.
Real-world objective functions are typically high-dimensional, and a
natural question is how the dimensionality of the landscape affects
the escape dynamics. Note that a basin serves as an idealized model of trapping near a local minimum in an optimization landscape.

For the cubic basins studied here, higher fixed dimension reduces
the asymptotic escape-time order as the basin depth grows.
Although a $D$-dimensional cubic basin of radius $N$ contains far more
states than its one-dimensional counterpart, the number of escape routes across shells near the boundary grows as $N^{D-1}$, providing the
polynomial factor in the escape-time scaling.
The shell structure of the infinity-norm basin
allows an exactly defined one-dimensional birth-death quotient chain
to be constructed from its electrical network, making the quotient analysis tractable.
For $D\ge2$, the shell index of the true walk is generally
not Markov. Nevertheless, shorting the shells yields a
resistance estimate that gives an order-matching lower bound
on the true walk's mean escape time, while an explicit flow
construction gives a matching upper bound.

We study a Metropolis random walk on the integer lattice $\Z^D$ with
energy $V(x)=\norm{x}_\infty$.
At each step, the walker proposes one of its $2D$ nearest-neighbor moves
uniformly; downhill moves are always accepted, while uphill moves are
accepted with probability $p=e^{-1/T}$.
We derive exact shell-averaged proposal probabilities $u_k$ and $v_k$
using a binomial identity, and show that the resulting birth-death chain
on the shell index admits a closed-form solution via a telescoping-product
technique.

The key contributions of this work are:
\begin{enumerate}
\item We derive an exact closed-form expression for the mean escape time
$Q_D(N)$ of the shell-quotient Metropolis chain on a $D$-dimensional
cubic basin of depth $N$ (Theorem~\ref{thm:exact-escape}).
\item We prove that the asymptotic scaling is
$\Theta(\alpha^N / N^{D-1})$ for $0<p<1$ (Theorem~\ref{thm:asymptotic}),
showing an asymptotic factor of order $1/N$ per additional fixed
dimension, up to constants depending on dimension and $p$.
\item We establish that the shell-averaged chain is exactly the quotient
chain obtained by shorting each shell in the true Metropolis network
(Proposition~\ref{prop:quotient}), and use Rayleigh monotonicity to prove
a rigorous lower bound of $\Omega(\alpha^N/N^{D-1})$ for the true chain
(Theorem~\ref{thm:lower-bound}).
\item We prove that the true Metropolis escape time has the same
$\Theta(\alpha^N/N^{D-1})$ scaling (Theorem~\ref{thm:true-scaling}),
by constructing an explicit low-energy flow to a face of the target
shell and applying Thomson's principle to get an order-matching upper bound.
\item We compare the analytical results against numerical solutions of the finite-state Metropolis chain for $D \in \{1, 2, 3\}$.
\end{enumerate}

The remainder of this paper is organized as follows.
Section~II reviews related work.
Section~III introduces the system model, including the $D$-dimensional
cubic basin, the Metropolis random walk, and the shell-averaged proposal
probabilities.
Section~IV derives the exact escape-time formula.
Section~V establishes the shell-quotient network construction, proves the asymptotic
scaling, and proves the full $\Theta(\alpha^N/N^{D-1})$ law for the true chain.
Section~VI presents numerical results.
Section~VII concludes with a discussion of implications and future directions.

\section{Related Work}

The theoretical analysis of SA has progressed from asymptotic convergence
guarantees to more practical finite-time characterizations.
We organize the relevant literature into several threads.

\textbf{Origins of simulated annealing.}
Kirkpatrick, Gelatt, and Vecchi~\cite{kirkpatrick1983} introduced simulated
annealing as an optimization heuristic inspired by the Metropolis
algorithm~\cite{metropolis1953} from statistical mechanics.
Geman and Geman~\cite{geman1984} established global convergence under
logarithmic cooling, and Hajek~\cite{hajek1988} clarified that schedules
must effectively be slower than the escape time from the deepest local minimum.
Catoni~\cite{catoni1992} developed rough large deviation estimates for
simulated annealing with application to exponential schedules.
These prior results establish that SA converges to the global optimum
under appropriate cooling, but they do not provide tractable expressions for
finite-time behavior at a fixed temperature.

\textbf{Finite-time and spectral methods.}
Desai and Rao~\cite{desai1997} studied the finite-time behavior of slowly
cooled annealing chains.
Holley, Kusuoka, and Stroock~\cite{holley1989} analyzed asymptotics of the
spectral gap with applications to the theory of simulated annealing.
Mitra, Romeo, and Sangiovanni-Vincentelli~\cite{mitra1986} studied convergence
and finite-time behavior of simulated annealing, establishing bounds on the
probability of finding optimal solutions within a given number of steps.
While these approaches yield rigorous general bounds, they often take the form
of inequalities rather than exact expressions.

\textbf{The one-dimensional predecessor.}
In~\cite{ramachandran2026}, we adopted the notion of basin depth and focused
on finite-time, constant-temperature regimes, quantifying the expected time to
escape.
We derived exact closed-form expressions for the mean escape time from a single
linear basin in one dimension, and extended the analysis to a two-basin landscape
to obtain exact expressions for the expected time to reach the global optimum.
We also demonstrated how the predicted escape time can guide the design of a
two-temperature switching strategy.
 The present paper extends that single-basin setting to cubic basins in any fixed dimension $D$: it gives an exact shell-quotient formula and a matching-order theorem for the true walk.

\textbf{Potential-theoretic metastability.}
Bovier, Eckhoff, Gayrard, and Klein~\cite{bovier2002} obtain precise
mean-exit-time and spectral results for a class of reversible discrete-state
Markov chains under metastability assumptions.
Bovier and den Hollander~\cite{bovier2015} develop the potential-theoretic
framework, relating hitting probabilities and times to capacities and
variational principles, with applications beyond one dimension.
Within this potential-theoretic framework, our contribution
is an explicit calculation for cubic basins in the
growing-depth limit $N\to\infty$ at fixed $D$ and $p$:
an exact shell-quotient formula and a constructive
matching-order resistance analysis that identifies the
factor $N^{-(D-1)}$.

\textbf{Exact and approximate state aggregation.}
Kemeny and Snell~\cite[Sec.~6.3, Thm.~6.3.2]{kemeny1976}
show that grouping states preserves the Markov dynamics when all states in each group have the same total transition probability to every other group. Grouping all points in a shell generally fails this condition for $D\ge2$, as illustrated in Section~III. For the numerical computations in Section~VI, we instead group
points that differ only by coordinate signs or coordinate order. These points are equivalent by symmetry, so this smaller numerical model preserves the true walk's escape time exactly.

\textbf{Random walks and first-passage times.}
Palyulin and Metzler~\cite{palyulin2012} studied how a finite potential barrier
affects the mean first-passage time, a problem closely related to our
barrier-crossing framework.
Roberts, Gelman, and Gilks~\cite{roberts1997} analyzed diffusion limits and
optimal scaling of random walk Metropolis algorithms, providing connections
between discrete Metropolis chains and continuous diffusion processes that
motivate the discretization corrections studied in~\cite{ramachandran2026}.

\textbf{Geometry in high-dimensional optimization.}
The idea that high-dimensional geometry can accelerate escape from
local minima has appeared in several contexts.
Dauphin et al.~\cite{dauphin2014} argue that saddle points
and their surrounding plateaus, rather than high-error local
minima, can be a major obstacle in high-dimensional
non-convex optimization.
In combinatorial optimization, Bellitti, Ricci-Tersenghi, and
Scardicchio~\cite{bellitti2021} demonstrated that for random constraint
satisfaction problems, the dominant obstacle for both classical and
quantum algorithms can be \emph{entropic} rather than energetic. In biophysics, Bicout and Szabo~\cite{bicout2000} study diffusion in
a spherically symmetric model of protein folding, illustrating entropic
barriers and folding kinetics. These related geometric considerations
motivate the present analysis.

\textbf{Electrical network methods.}
The Rayleigh monotonicity principle and Thomson's variational principle for
electrical networks provide powerful tools for bounding hitting times of
reversible Markov chains~\cite{doyle1984,levin2017,tetali1991}.
We use these tools in Sections~\ref{sec:lower-bound}
and~\ref{sec:upper-bound} to establish the full
scaling law for the true Metropolis chain.
Rayleigh monotonicity (shorting nodes can only decrease resistance) gives
the lower bound, while Thomson's principle (any unit flow upper-bounds the
resistance) gives the matching upper bound via an explicit flow construction.

\section{System Model}

\subsection{The $D$-Dimensional Cubic Basin}

Consider a $D$-dimensional cubic basin on the integer lattice $\Z^D$
with energy function
\begin{equation}
\label{eq:energy}
V(x) = \norm{x}_\infty = \max(|x_1|,\dots,|x_D|).
\end{equation}
The minimum-energy state is the origin, and the energy increases
by one unit with each concentric shell.
We define the \emph{shell} at level $k$ as
\[
A_k := \{x \in \Z^D : \norm{x}_\infty = k\}, \quad k=0,1,\dots,N,
\]
where $N$ is the basin depth.
Shell $N$ is treated as absorbing: once the walker reaches any state
with $\norm{x}_\infty = N$, the escape is complete.

For a site on shell $k \ge 1$, let $s$ denote the number of coordinates
with $|x_i|=k$; we call these the \emph{boundary coordinates}.
The number of sites with exactly $s$ boundary coordinates on shell $k$ is
\begin{equation}
\label{eq:MDs}
M_{D,s}(k) = \binom{D}{s} 2^s (2k-1)^{D-s}, \quad s=1,\dots,D.
\end{equation}
There are $\binom{D}{s}$ ways to choose which $s$ coordinates are at $\pm k$,
$2^s$ sign choices for those coordinates, and $(2k-1)^{D-s}$ values
in $\{-(k-1),\dots,k-1\}$ for the remaining coordinates.
The total number of sites on shell $k$ is therefore
\begin{equation}
\label{eq:SD}
S_D(k) = \sum_{s=1}^{D} M_{D,s}(k) = (2k+1)^D - (2k-1)^D.
\end{equation}
The identity \eqref{eq:SD} follows from the binomial theorem: $(2k+1)^D$
counts all integer points in $\{-k,\dots,k\}^D$, and $(2k-1)^D$ counts
those strictly inside.
We use $S_D(0):=1$ separately; formula~\eqref{eq:SD} applies for $k\ge1$.

\begin{figure}[!t]
\centering
\includegraphics[width=\columnwidth]{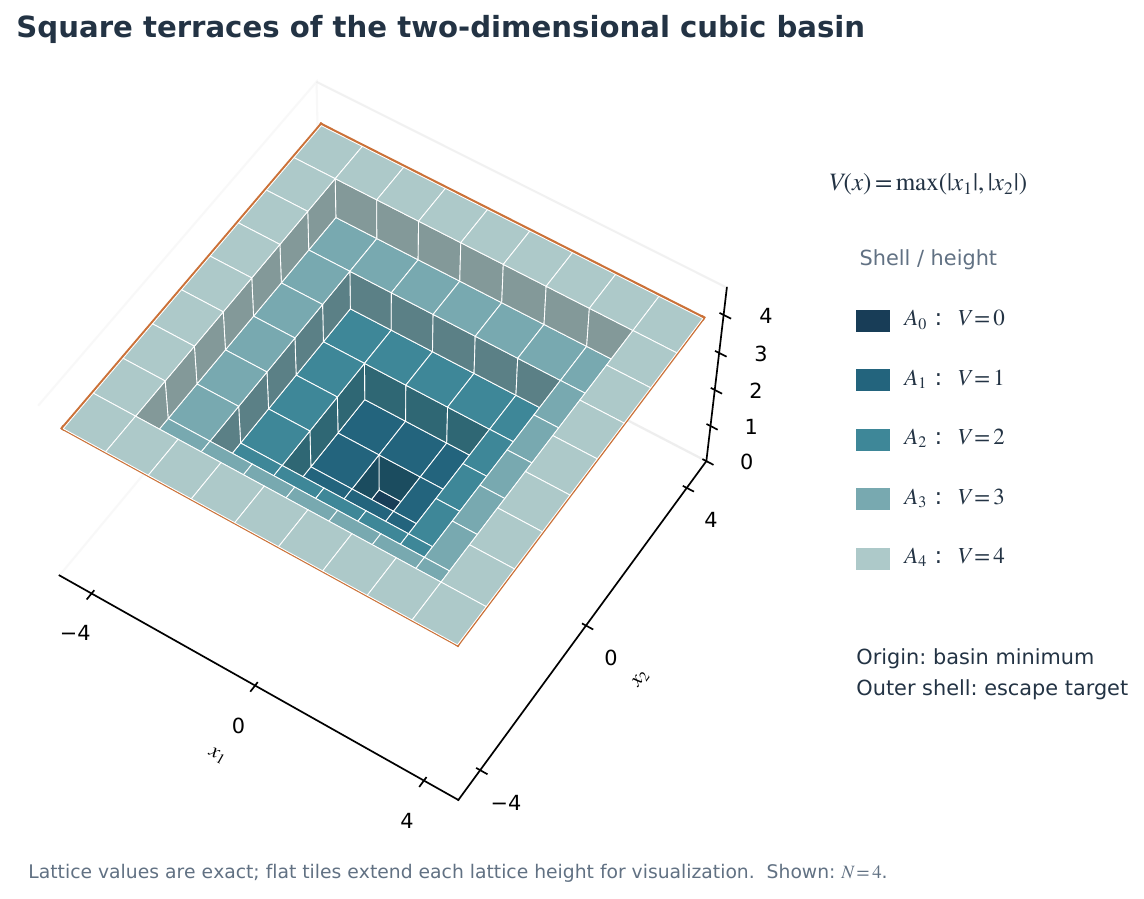}
\caption{The cubic basin for $D=2$ and $N=4$.
Each square terrace represents a shell $A_k$ at energy
$V(x)=\norm{x}_\infty=k$. The origin is the basin minimum;
escape occurs upon reaching the outermost shell $A_N$.
The flat tiles extend the discrete lattice heights to
visualize the stepped geometry.}
\label{fig:cubic-basin}
\end{figure}

Figure~\ref{fig:cubic-basin} illustrates the geometry for $D=2$:
the shells form concentric square terraces, with each outward
step increasing the energy by one.

\subsection{The Metropolis Random Walk}

We study a fixed-temperature Metropolis random walk on $\Z^D$
with energy~\eqref{eq:energy}.
At each time step, the walker at state $x$ proposes one of the $2D$
nearest-neighbor moves $x \pm e_i$ uniformly at random, where $e_i$
is the $i$-th standard basis vector.
If the proposed move decreases the energy ($V(y) < V(x)$), it is always
accepted.
If the move increases the energy ($V(y) > V(x)$), it is accepted with
probability
\begin{equation}
\label{eq:p-def}
p = e^{-\Delta E / T} = e^{-1/T},
\end{equation}
since each uphill step increases $V$ by exactly one unit.
Moves that keep the energy unchanged are accepted with probability~$1$.

We write $\alpha := 1/p\ge1$ for the \emph{bias ratio};
$\alpha>1$ when $p<1$, governing the exponential barrier to escape.
The first-passage time is
\begin{equation}
\label{eq:tau}
\tau_N := \inf\{t \ge 0 : \norm{X_t}_\infty = N\},
\end{equation}
starting from the origin $X_0 = 0$.
We distinguish the true-chain mean escape time $H_D(N):=\E_0[\tau_N]$
from the shell-quotient mean escape time $Q_D(N)$ defined below.
The dependence on $p$ is suppressed in both notations.

\subsection{Shell-Averaged Proposal Probabilities}

We define a simplified chain on the shell indices by averaging
the outward and inward proposal probabilities uniformly over
all sites in each shell. For $D\ge2$, these probabilities vary
within a shell, so the resulting shell-quotient chain does not
exactly describe the true walk's shell transitions. For example,
at $(k,0,\ldots,0)$ only one proposal direction leads outward,
whereas at $(k,k,0,\ldots,0)$ two do, giving outward proposal
probabilities $1/(2D)$ and $1/D$, respectively, for $1\le k<N$.

A site with $s$ boundary coordinates has $s$ outward directions (each
increasing $\norm{x}_\infty$ from $k$ to $k+1$) out of $2D$ total
directions.
The outward proposal probability averaged over shell $k$ is
\begin{equation}
\label{eq:uk-sum}
u_k = \frac{1}{2D\,S_D(k)} \sum_{s=1}^{D} s\,M_{D,s}(k).
\end{equation}
Substituting~\eqref{eq:MDs} gives
$u_k = \frac{1}{2D\,S_D(k)} \sum_{s=1}^{D} s\binom{D}{s}2^s(2k-1)^{D-s}$.
We simplify this using the binomial identity
\begin{equation}
\label{eq:binomial-identity}
\sum_{s=1}^{D} s\binom{D}{s}2^s a^{D-s} = 2D(a+2)^{D-1},
\end{equation}
which can be derived by differentiating $(a+2)^D = \sum_{s=0}^{D}
\binom{D}{s}2^s a^{D-s}$ with respect to $a$, multiplying by $a$,
and subtracting from $D(a+2)^D$.
Setting $a = 2k-1$ yields $\sum_{s=1}^{D} s\,M_{D,s}(k) = 2D(2k+1)^{D-1}$,
and therefore
\begin{equation}
\label{eq:uk}
\boxed{u_k = \frac{(2k+1)^{D-1}}{S_D(k)}.}
\end{equation}

For the inward direction, only sites with exactly $s=1$ boundary coordinate
can step inward: if $s \ge 2$, reducing the absolute value of one boundary coordinate from $k$ to $k-1$ leaves $\norm{x}_\infty = k$ unchanged.
Each such site has one inward direction out of $2D$, and the number of
such sites is $M_{D,1}(k) = 2D(2k-1)^{D-1}$.
Therefore
\begin{equation}
\label{eq:vk}
\boxed{v_k = \frac{(2k-1)^{D-1}}{S_D(k)}.}
\end{equation}

Note that $u_k$ and $v_k$ are \emph{proposal} probabilities.
Since outward moves are accepted with probability $p$ and inward moves
are always accepted, the actual shell-averaged transition probabilities
are $p\,u_k$ (outward) and $v_k$ (inward).
The remaining probability $r_k = 1 - p\,u_k - v_k$ accounts for
same-shell moves (both accepted lateral proposals and rejected uphill
proposals).

Intuitively, for large $k$, most sites on a shell have exactly one boundary
coordinate, giving one outward and one inward proposal direction.
Accordingly, both $u_k$ and $v_k$ approach $1/(2D)$ as
$k\to\infty$. The corresponding outward transition probability
approaches $p/(2D)$.

\section{Exact Escape Time}

\subsection{Birth-Death Recurrence}

Let $T_k$ denote the mean first-passage time from state $k$ to the absorbing
state $N$ of the shell-quotient birth-death chain.
We retain $T_k$ in this derivation and write $Q_D(N):=T_0^{(D)}(N)$
for its escape time from state $0$. All recurrences below with $k\ge1$ apply to transient indices
$1\le k<N$, with boundary condition $T_N=0$.
The first-step equation for $k \ge 1$ is
\[
T_k = 1 + p\,u_k\,T_{k+1} + v_k\,T_{k-1} + r_k\,T_k.
\]
Substituting $r_k = 1 - p\,u_k - v_k$ and rearranging, we obtain
\begin{equation}
\label{eq:first-step}
p\,u_k\,(T_k - T_{k+1}) - v_k\,(T_{k-1} - T_k) = 1.
\end{equation}
We define the \emph{backward differences}
$b_k := T_k - T_{k+1}$,
so that~\eqref{eq:first-step} becomes
\begin{equation}
\label{eq:recurrence}
\boxed{p\,u_k\,b_k = v_k\,b_{k-1} + 1, \qquad k \ge 1.}
\end{equation}

At the origin ($k=0$), all $2D$ neighbors lie on shell $1$, and each
proposed move is uphill with acceptance probability $p$.
By symmetry, $T_0 = 1 + p\,T_1 + (1-p)\,T_0$, which gives
\begin{equation}
\label{eq:b0}
\boxed{b_0 := T_0 - T_1 = \frac{1}{p}.}
\end{equation}

We define the ratio
\begin{equation}
\label{eq:Rk}
R_k := \frac{v_k}{p\,u_k}
= \frac{1}{p}\left(\frac{2k-1}{2k+1}\right)^{D-1},
\end{equation}
so the recurrence can be written as $b_k = R_k\,b_{k-1} + 1/(p\,u_k)$.

Unrolling the first few terms:
\begin{align*}
b_1 &= R_1\,b_0 + \frac{1}{p\,u_1},\\[3pt]
b_2 &= R_2 R_1\,b_0 + R_2\frac{1}{p\,u_1} + \frac{1}{p\,u_2},\\[3pt]
b_3 &= R_3 R_2 R_1\,b_0 + R_3 R_2\frac{1}{p\,u_1}
  + R_3\frac{1}{p\,u_2} + \frac{1}{p\,u_3}.
\end{align*}
This suggests, and an induction argument confirms, the general form
\begin{equation}
\label{eq:bk-general}
b_k = \left(\prod_{j=1}^{k} R_j\right) b_0
+ \sum_{m=1}^{k} \frac{1}{p\,u_m}\prod_{j=m+1}^{k} R_j.
\end{equation}

\subsection{Telescoping and Closed Form}

The product of ratios telescopes.
From~\eqref{eq:Rk}, for $k \ge m+1$,
\[
\prod_{j=m+1}^{k} R_j
= \frac{1}{p^{k-m}} \prod_{j=m+1}^{k}
  \left(\frac{2j-1}{2j+1}\right)^{D-1}.
\]
The inner product telescopes:
\[
\prod_{j=m+1}^{k} \frac{2j-1}{2j+1}
= \frac{2m+1}{2m+3}\cdot\frac{2m+3}{2m+5}\cdots\frac{2k-1}{2k+1}
= \frac{2m+1}{2k+1}.
\]
Therefore
\begin{equation}
\label{eq:telescope}
\prod_{j=m+1}^{k} R_j
= \frac{1}{p^{k-m}}\left(\frac{2m+1}{2k+1}\right)^{D-1}.
\end{equation}

Substituting~\eqref{eq:telescope} into~\eqref{eq:bk-general} with
$b_0 = 1/p$ and $1/(p\,u_m) = S_D(m)/\bigl(p(2m+1)^{D-1}\bigr)$:

\emph{The $m=0$ term} (from $b_0$):
\[
\left(\prod_{j=1}^{k} R_j\right)\frac{1}{p}
= \frac{1}{p^{k+1}(2k+1)^{D-1}}.
\]
\emph{The $m \ge 1$ terms}: the factors $(2m+1)^{D-1}$ cancel between
$1/(p\,u_m)$ and the telescoping product, giving
\[
\frac{1}{p\,u_m}\prod_{j=m+1}^{k} R_j
= \frac{S_D(m)}{p^{k-m+1}(2k+1)^{D-1}}.
\]
Combining and writing $\alpha = 1/p$, we obtain
\begin{equation}
\label{eq:bk-closed}
\boxed{
b_k = \frac{1}{(2k+1)^{D-1}}
\left[\alpha^{k+1} + \sum_{m=1}^{k} S_D(m)\,\alpha^{k-m+1}\right].
}
\end{equation}

Since $T_N = 0$ and $b_\ell = T_\ell - T_{\ell+1}$, we have
$T_0 = \sum_{\ell=0}^{N-1} b_\ell$.
This yields our first main result.

\begin{theorem}[Exact escape time]
\label{thm:exact-escape}
For the shell-averaged Metropolis random walk on a $D$-dimensional cubic
basin of depth $N$ with acceptance probability $p \in (0,1]$ and
$\alpha = 1/p$, the mean escape time from the origin is
{\small
\begin{equation}
\label{eq:T0-exact}
\boxed{
T_0^{(D)}(N) = \sum_{\ell=0}^{N-1}
\frac{1}{(2\ell+1)^{D-1}}
\left[\alpha^{\ell+1}
+ \sum_{m=1}^{\ell} S_D(m)\,\alpha^{\ell-m+1}\right],
}
\end{equation}
\par}
where $S_D(m) = (2m+1)^D - (2m-1)^D$ is the number of lattice points on
shell~$m$.
Thus this formula gives $Q_D(N)$ exactly. It also gives $H_1(N)$
when $D=1$, but is used as an approximation to $H_D(N)$ for $D\ge2$.
\end{theorem}

Each term $b_\ell$ in the sum can be interpreted as follows: the factor
$\alpha^{\ell+1}$ captures the exponential cost of climbing from shell
$\ell$ to shell $\ell+1$, the shell-size terms $S_D(m)$ weight the
contributions of intermediate shells, and the prefactor
$1/(2\ell+1)^{D-1}$ captures the speedup with higher dimensions: the larger the
shell surface, the easier the outward step.

\subsection{Special Cases}

We verify Theorem~\ref{thm:exact-escape} by examining low-dimensional
special cases.

\textbf{$D=1$:} Here $S_1(m) = 2$ and $(2\ell+1)^{D-1}=1$, so
\begin{equation}
\label{eq:T0-1D}
T_0^{(1)} = \sum_{\ell=0}^{N-1}
\left[\alpha^{\ell+1} + 2\sum_{m=1}^{\ell} \alpha^{\ell-m+1}\right].
\end{equation}

\begin{remark}[Recovery of the one-dimensional result]
\label{rmk:1D}
Formula~\eqref{eq:T0-1D} coincides with the single-basin escape
time derived in~\cite{ramachandran2026}, after identifying our acceptance
probability $p$ with $2p_{\mathrm{prior}}$, where the factor of two accounts
for the fact that~\cite{ramachandran2026} absorbs the one-dimensional
proposal probability $1/2$ into the definition of~$p$.
In both cases, $\alpha = e^{1/T}$ represents the same physical bias ratio.
\end{remark}

\textbf{$D=2$:} Here $S_2(m) = (2m+1)^2 - (2m-1)^2 = 8m$, and the
formula becomes
\begin{equation}
\label{eq:T0-2D}
T_0^{(2)} = \sum_{\ell=0}^{N-1}
\frac{1}{2\ell+1}
\left[\alpha^{\ell+1} + \sum_{m=1}^{\ell} 8m\,\alpha^{\ell-m+1}\right].
\end{equation}
The $1/(2\ell+1)$ prefactor produces the $1/N$ asymptotic speedup relative
to one dimension.

\textbf{$D=3$:} Here $S_3(m) = (2m+1)^3 - (2m-1)^3 = 24m^2+2$, giving
\begin{equation}
\label{eq:T0-3D}
T_0^{(3)} = \sum_{\ell=0}^{N-1}
\frac{1}{(2\ell+1)^2}
\left[\alpha^{\ell+1}
+ \sum_{m=1}^{\ell} (24m^2+2)\,\alpha^{\ell-m+1}\right].
\end{equation}

\begin{remark}[The unbiased case $p=1$]
\label{rmk:unbiased}
When $p=1$ (so $\alpha=1$), the Cartesian random walk is unbiased.
The shell sizes telescope: $1 + \sum_{m=1}^{\ell}S_D(m) = (2\ell+1)^D$,
so $b_\ell = (2\ell+1)^D / (2\ell+1)^{D-1} = 2\ell+1$.
Thus $Q_D(N)=\sum_{\ell=0}^{N-1}(2\ell+1)=N^2$ for every
$D$ when $p=1$. This dimension-independent identity applies
to the quotient chain; the true-chain mean $H_D(N)$ need
not equal $N^2$.
\end{remark}

\section{Shell-Quotient Construction and Asymptotic Scaling}

\subsection{Construction of the Birth-Death Quotient}
\label{sec:quotient}

\begin{figure}[!t]
\centering
\includegraphics[width=\columnwidth]{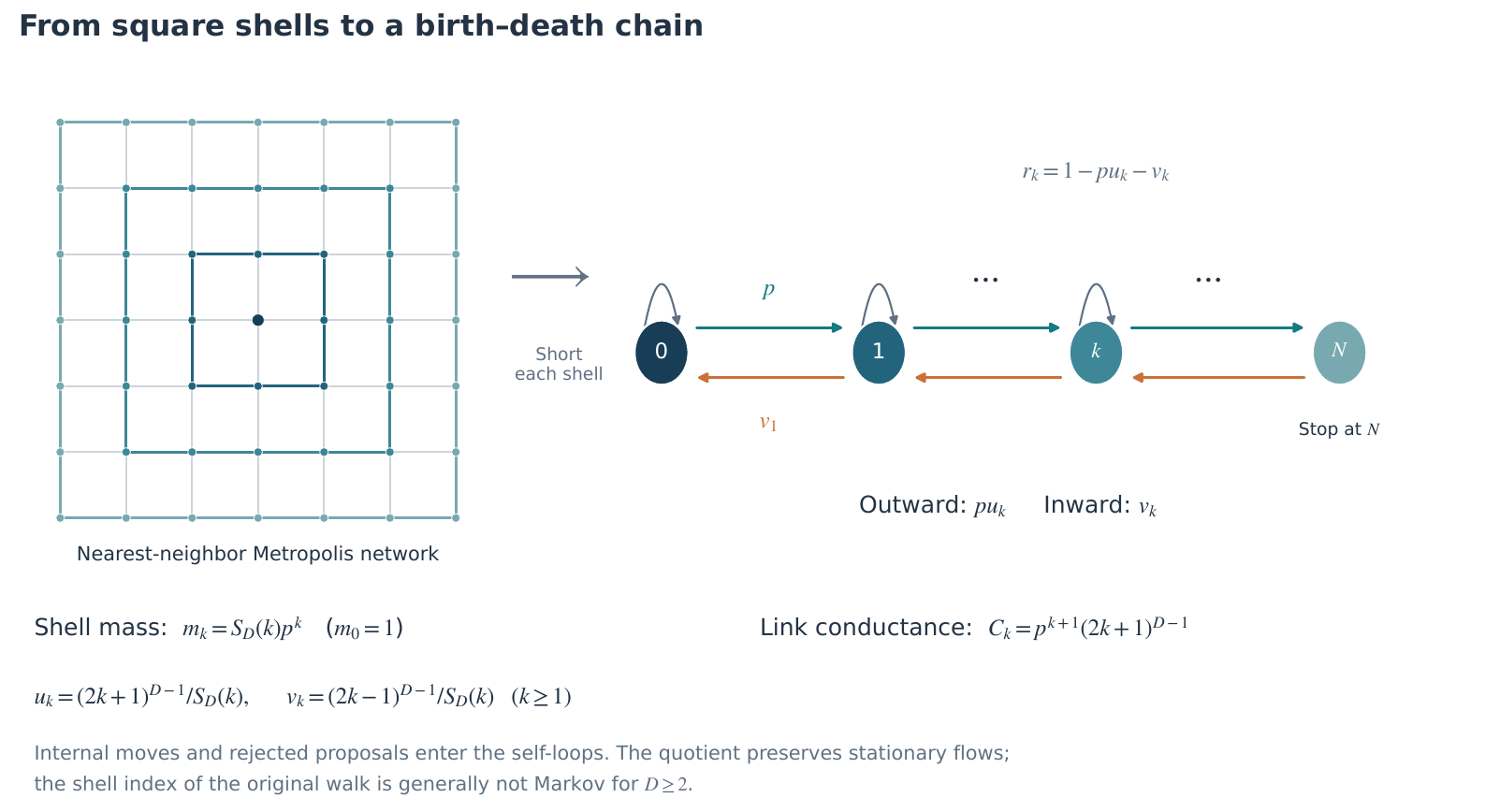}
\caption{Shell-quotient construction, illustrated for $D=2$.
Shorting each shell produces a birth--death chain with
shell mass $m_k=S_D(k)p^k$ ($m_0=1$) and link conductance
$C_k=p^{k+1}(2k+1)^{D-1}$.
For $1\le k<N$, outward and inward transitions have
probabilities $pu_k$ and $v_k$; internal moves and rejected
proposals contribute to self-loops. Escape calculations stop
at the first visit to $N$.}
\label{fig:shell-quotient}
\end{figure}

The shell-averaged chain is obtained by collapsing each shell
in the true Metropolis electrical network to a single node,
as illustrated in Figure~\ref{fig:shell-quotient}.
The node retains the shell's total stationary mass, while
conductances between successive shells are added.
Transitions within a shell become self-loops.

\begin{proposition}[The shell-averaged chain is the quotient chain]
\label{prop:quotient}
Fix $N$ and consider the true Metropolis chain restricted to
$B_N := \{x \in \Z^D : \norm{x}_\infty \le N\}$,
 with any proposal leaving $B_N$ rejected. This un-killed, reflecting
chain is reversible with respect to
\begin{equation}
\label{eq:pi}
\pi(x) \propto p^{\norm{x}_\infty}.
\end{equation}
 Short each shell $A_k$ to a single node and retain its total stationary
mass. For $1\le k<N$, the quotient birth-death chain has transition
probabilities $p\,u_k$ (outward) and $v_k$ (inward); at $k=0$ its
outward probability is $p$. For escape-time calculations, stop either
chain at its first visit to shell $N$. 
\end{proposition}

\begin{proof}
Reversibility follows from detailed balance.
If $x \sim y$ and $V(y) = V(x)+1$, then
\[
\pi(x)\,P(x,y) \propto p^{V(x)}\cdot\frac{p}{2D}
= \frac{p^{V(y)}}{2D}
\propto \pi(y)\,P(y,x),
\]
since $P(y,x) = 1/(2D)$ for downhill moves.
Equal-energy edges have symmetric proposal probabilities and equal
weights, and rejected proposals contribute only self-loops. For the
network calculation we use the unnormalized measure
$\mu(x)=p^{\norm{x}_\infty}$, so conductances below are exactly
$c(x,y)=\mu(x)P(x,y)$, without a normalization factor.

The total stationary mass of shell $A_k$ is proportional to $S_D(k)\,p^k$.
In this normalization the shell mass is $S_D(k)p^k$, with $S_D(0)=1$.
We count the edges crossing from shell $k$ to shell $k+1$.
From the shell-counting identity~\eqref{eq:binomial-identity},
the total number of oriented outward edges is $2D(2k+1)^{D-1}$.
Each such edge carries conductance $p^{k+1}/(2D)$ (from $\pi(x)\,P(x,y)$
with $V(x)=k$), so the total conductance across the cut
$A_k \leftrightarrow A_{k+1}$ is
\begin{equation}
\label{eq:conductance}
C_k = p^{k+1}(2k+1)^{D-1}.
\end{equation}
The quotient chain therefore has outward transition probability
\[
\widetilde{P}(k,k+1) = \frac{C_k}{S_D(k)\,p^k}
= p\,\frac{(2k+1)^{D-1}}{S_D(k)} = p\,u_k,
\]
and inward transition probability
{\small
\[
\widetilde{P}(k,k-1) = \frac{C_{k-1}}{S_D(k)\,p^k}
= \frac{p^k(2k-1)^{D-1}}{S_D(k)\,p^k}
= \frac{(2k-1)^{D-1}}{S_D(k)} = v_k.
\]
\par}
These are exactly the shell-averaged probabilities, completing the proof.
The displayed inward and outward formulas apply for $1\le k<N$.
At $k=0$, the $2D$ edges to $A_1$ each have conductance $p/(2D)$,
so $C_0=p$ and $\widetilde P(0,1)=p$. Internal transitions and rejected
proposals contribute to the quotient self-loop, preserving the shell mass.
\end{proof}


\subsection{Asymptotic Scaling}

\begin{theorem}[Asymptotic scaling]
\label{thm:asymptotic}
Fix $D \ge 1$ and $0<p<1$, and let $\alpha = 1/p > 1$.
Then the shell-averaged escape time satisfies
$Q_D(N)=T_0^{(D)}(N)$ and
\begin{equation}
\label{eq:theta-scaling}
T_0^{(D)}(N) = \Theta\!\left(\frac{\alpha^N}{N^{D-1}}\right)
\qquad (N \to \infty).
\end{equation}
In particular,
$T_0^{(1)}(N) = \Theta(\alpha^N)$,
$T_0^{(2)}(N) = \Theta(\alpha^N/N)$, and
$T_0^{(3)}(N) = \Theta(\alpha^N/N^2)$.
\end{theorem}

\begin{proof}
From~\eqref{eq:bk-closed}, factor out $\alpha^{\ell+1}$:
\begin{equation}
\label{eq:bell-factored}
b_\ell = \frac{\alpha^{\ell+1}}{(2\ell+1)^{D-1}}\,F_\ell,
\end{equation}
where
$F_\ell := 1 + \sum_{m=1}^{\ell} S_D(m)\,\alpha^{-m}$.
Since $S_D(m) = \Theta(m^{D-1})$ and $\alpha > 1$, the series
$\sum_{m \ge 1} S_D(m)\,\alpha^{-m}$
converges.
Hence $F_\ell$ is increasing and bounded:
$1 \le F_\ell \le F_\infty < \infty$, where
$F_\infty = 1 + \sum_{m \ge 1} S_D(m)\,\alpha^{-m}$.
It follows that $b_\ell \asymp \alpha^\ell / \ell^{D-1}$.

Next, the ratio of consecutive terms satisfies
\[
\frac{b_{\ell-1}}{b_\ell}
= \alpha^{-1}\left(\frac{2\ell+1}{2\ell-1}\right)^{D-1}
  \frac{F_{\ell-1}}{F_\ell}
\longrightarrow \alpha^{-1} < 1.
\]
Hence there exist $q \in (0,1)$ and $L$ such that
$b_{\ell-1} \le q\,b_\ell$ for all $\ell \ge L$.
The sum $T_0^{(D)}(N) = \sum_{\ell=0}^{N-1} b_\ell$ is therefore
geometrically dominated by its last term:
\[
b_{N-1} \le T_0^{(D)}(N)
\le \sum_{\ell=0}^{L-1} b_\ell + \frac{1}{1-q}\,b_{N-1}
\le C'\,b_{N-1}.
\]
Since $b_{N-1} = \Theta(\alpha^N / N^{D-1})$, the claim follows.
\end{proof}

\begin{remark}[Sharper asymptotic]
\label{rmk:sharp}
The proof also yields the sharper estimate
\begin{equation}
\label{eq:sharp-asymptotic}
T_0^{(D)}(N) \sim \frac{\alpha}{\alpha-1}\,F_\infty\,
\frac{\alpha^N}{(2N)^{D-1}}.
\end{equation}
To justify the leading constant, put $r:=\alpha^{-1}<1$.
For each fixed $j\ge0$, the ratio limit in the proof gives
$b_{N-1-j}/b_{N-1}\to r^j$ as $N\to\infty$.
Choose $r<q<1$ and $L$ such that $b_{\ell-1}\le q b_\ell$
for all $\ell\ge L$.
The terms with $\ell\ge L$, read backward from $N-1$ and
divided by $b_{N-1}$, are bounded by the summable sequence $q^j$.
The fixed prefix with $\ell<L$ contributes a quantity tending
to zero, since $b_{N-1}\to\infty$.
Dominated convergence therefore gives
\[
\frac{T_0^{(D)}(N)}{b_{N-1}}
\longrightarrow \sum_{j=0}^{\infty}r^j
=\frac{\alpha}{\alpha-1}.
\]
Combining this with
$b_{N-1}=\alpha^N F_{N-1}/(2N-1)^{D-1}$ and
$F_{N-1}\to F_\infty$ proves the stated equivalent.

\end{remark}

The scaling $\Theta(\alpha^N/N^{D-1})$ shows that escape time
grows exponentially with $N$, with the base $\alpha$ determined
by temperature. Higher dimension reduces this time by a
polynomial factor because the outermost shells provide
$\Theta(N^{D-1})$ escape routes.




\subsection{Hitting-Time Identity}
\label{sec:hitting-time}
Throughout the resistance and hitting-time arguments, $D\ge1$ and
$0<p<1$ are fixed. Unless a finite restriction is stated explicitly,
let $P(x,y)$ denote the one-step transition probability
from $x$ to $y$ for the unkilled Metropolis chain on $\Z^D$, stopped only when
its specified target is reached. Constants may depend on $D$ and $p$,
but not on $N$.

To relate escape time to effective resistance, we use
the reversible measure $\mu(x) := p^{\norm{x}_\infty}$, under which
the chain satisfies detailed balance:
$\mu(x)\,P(x,y) = \mu(y)\,P(y,x)$ for all adjacent $x,y$.
The associated edge conductances are $c(x,y) = \mu(x)\,P(x,y)$,
and the Dirichlet form is
\[
\Energy(f,f) := \tfrac{1}{2}\sum_{x \sim y} c(x,y)\bigl(f(x)-f(y)\bigr)^2.
\]
Here $\sum_{x\sim y}$ counts both orientations of each distinct
nearest-neighbor edge. Equivalently, the Dirichlet form sums once over
unoriented edges without the factor $1/2$. All flow-energy sums below
count each unoriented edge once, using any fixed orientation and an
antisymmetric flow $\theta(y,x)=-\theta(x,y)$:
\[
\operatorname{div}\theta(x):=\sum_{y\sim x}\theta(x,y),
\qquad \mathcal E_{\rm flow}(\theta):=
\sum_{\{x,y\}}\frac{\theta(x,y)^2}{c(x,y)}.
\]
Self-loops do not contribute to either energy; their contribution to
the holding times is retained through $\mu$ and $P$.
The capacity between a point $a$ and a set $A$ is
$\mathrm{Cap}(a,A) := \inf\{\Energy(f,f) : f(a)=1,\, f|_A=0\}$.
By Dirichlet's principle~\cite{doyle1984}, this minimum is
the effective conductance between $a$ and $A$. Hence
$R_{\mathrm{eff}}(a\leftrightarrow A)=1/\mathrm{Cap}(a,A)$.

\begin{lemma}[Hitting time via capacity]
\label{lem:hitting}
 For a nonempty finite target $A\subset\Z^D$ and $a\notin A$, let
$\tau_A:=\inf\{t\ge0:X_t\in A\}$, $\tau_a:=\tau_{\{a\}}$, and
$\tau_a^+:=\inf\{t\ge1:X_t=a\}$. Then
\begin{equation}
\label{eq:hitting-identity}
\E_a[\tau_A] = \frac{1}{\mathrm{Cap}(a,A)}
\sum_{x} \mu(x)\,\mathrm{Pr}_x(\tau_a < \tau_A).
\end{equation}
In particular,
\begin{equation}
\label{eq:hitting-bounds}
\mu(a)\,R_{\mathrm{eff}}(a \leftrightarrow A)
\le \E_a[\tau_A]
\le M_{D,p}\,R_{\mathrm{eff}}(a \leftrightarrow A),
\end{equation}
where $M_{D,p} := \sum_{x \in \Z^D} \mu(x) < \infty$.
\end{lemma}

\begin{proof}
The reversible measure $\mu(x)=p^{\norm{x}_\infty}$ has finite
total mass
\[
M_{D,p}=1+\sum_{k\ge1}S_D(k)p^k<\infty,
\]
because $S_D(k)=O(k^{D-1})$ and $p<1$.
Thus the irreducible chain has stationary distribution
$\mu/M_{D,p}$ and is positive recurrent. Consequently, the
hitting and return times considered below have finite means.

We first establish the capacity identity. Put
$h(x)=\Pr_x(\tau_a<\tau_A)$, the probability of reaching $a$
before $A$. This function equals $1$ at $a$ and $0$ on $A$.
At every other state, conditioning on the first step gives
\[
h(x)=\sum_yP(x,y)h(y).
\]
For two functions, the corresponding mixed term is
\[
\mathcal E(f,g)
=\frac12\sum_{x\sim y}c(x,y)
\bigl(f(x)-f(y)\bigr)\bigl(g(x)-g(y)\bigr).
\]
Expanding this sum and using reversibility,
$c(x,y)=c(y,x)$, yields
\[
\mathcal E(f,g)
=\sum_x\mu(x)g(x)
\left(f(x)-\sum_yP(x,y)f(y)\right).
\]
Self-loops contribute zero to the energy and may be included
when using $\sum_yP(x,y)=1$.

Taking $f=g=h$, only the term at $a$ remains: on $A$,
$h$ is zero, and elsewhere the expression in parentheses
is zero. Hence
\[
\begin{aligned}
\mathcal E(h,h)
&=\mu(a)\left(1-\sum_yP(a,y)h(y)\right)\\
&=\mu(a)\Pr_a(\tau_A<\tau_a^+).
\end{aligned}
\]
The last equality follows by conditioning on the first step:
$\sum_yP(a,y)h(y)$ is the probability of returning to $a$
before reaching $A$.

We now show that $h$ minimizes energy among functions equal
to $1$ at $a$ and $0$ on $A$. It suffices to consider functions
with values in $[0,1]$, since truncating to this interval
preserves the boundary values and cannot increase energy.
For any such function $f$, write $f=h+g$. Then $g=0$ on
$\{a\}\cup A$, so
\[
\mathcal E(h,g)
=\sum_x\mu(x)g(x)
\left(h(x)-\sum_yP(x,y)h(y)\right)=0.
\]
Indeed, $g$ vanishes on the boundary, and the expression in
parentheses vanishes elsewhere. Expanding the energy gives
\[
\mathcal E(f,f)
=\mathcal E(h,h)+2\mathcal E(h,g)+\mathcal E(g,g)
\ge\mathcal E(h,h).
\]
All these sums converge absolutely because the functions
are bounded and $\sum_x\mu(x)<\infty$. Thus $h$ attains
the infimum defining capacity, and
\[
\mathrm{Cap}(a,A)
=\mu(a)\Pr_a(\tau_A<\tau_a^+).
\]

To obtain the expected hitting time, let
\[
G_A(u,x)
:=\E_u\left[\sum_{t=0}^{\tau_A-1}
\mathbf1\{X_t=x\}\right]
\]
be the expected number of visits to $x$ before reaching $A$,
starting from $u$. Summing over visited states gives
\[
\E_a[\tau_A]=\sum_xG_A(a,x).
\]
Reversibility, applied to paths that avoid $A$, gives
\[
\mu(a)G_A(a,x)=\mu(x)G_A(x,a).
\]
Starting from $x$, visits to $a$ occur only if the chain
reaches $a$ before $A$. The strong Markov property therefore
gives
\[
G_A(x,a)=h(x)G_A(a,a).
\]
Starting from $a$, each visit to $a$ is followed by probability
$q=\Pr_a(\tau_A<\tau_a^+)$ of reaching $A$ before the next
return. Thus the number of visits to $a$, including the
initial visit, is geometric with mean $1/q$, so
\[
G_A(a,a)
=\frac{1}{q}
=\frac{\mu(a)}{\mathrm{Cap}(a,A)}.
\]
Combining these identities yields
\[
\begin{aligned}
\E_a[\tau_A]
&=\frac{\sum_x\mu(x)h(x)}{\mathrm{Cap}(a,A)}\\
&=R_{\mathrm{eff}}(a\leftrightarrow A)\sum_x\mu(x)h(x),
\end{aligned}
\]
which is~\eqref{eq:hitting-identity}.
Finally, $h(a)=1$ and $0\le h(x)\le1$ imply
\[
\mu(a)\le\sum_x\mu(x)h(x)\le M_{D,p},
\]
giving~\eqref{eq:hitting-bounds}.
\end{proof}

For the fixed starting point $a=0$, where $\mu(0)=1$, and fixed
$D,p$, Lemma~\ref{lem:hitting} compares the mean hitting time with
effective resistance using constants independent of the target depth
$N$. This enables both lower and upper bounds through resistance estimates.

\subsection{Lower Bound for the True Chain}
\label{sec:lower-bound}

\begin{figure*}[!t]
\centering
\includegraphics[width=\textwidth]{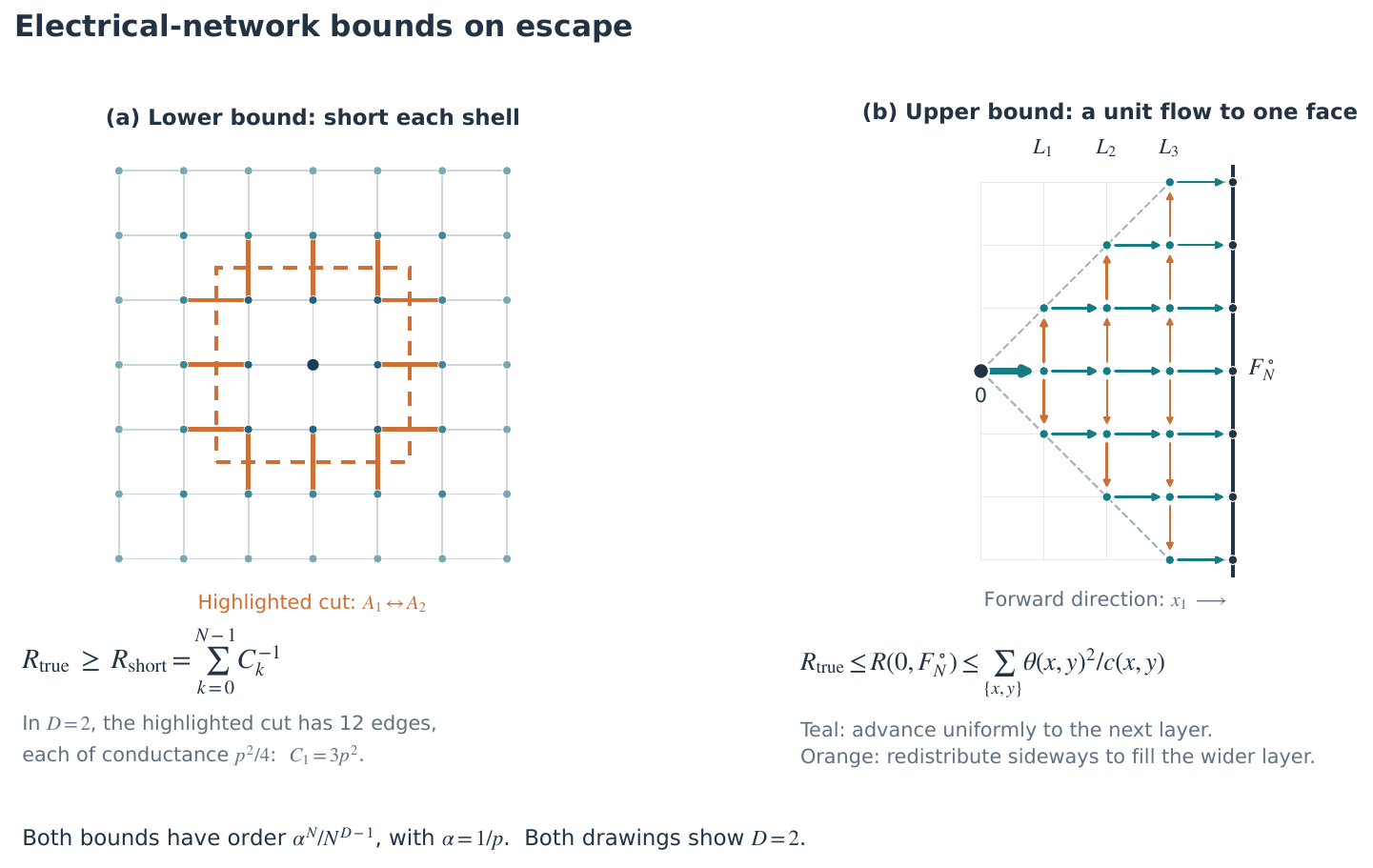}
\caption{Electrical-network bounds, illustrated for $D=2$.
(a) Shorting each shell gives the resistance lower bound
$R_{\mathrm{eff}}(0,A_N)\ge\sum_{k=0}^{N-1}C_k^{-1}$.
The highlighted cut between $A_1$ and $A_2$ contains
12 edges, each of conductance $p^2/4$, giving $C_1=3p^2$.
(b) A unit flow through an expanding pyramid subnetwork
gives the resistance upper bound by Thomson's principle.
Teal arrows carry current forward between layers;
orange arrows redistribute it within each wider layer.
The final step reaches the target face $F_N^\circ$.
The forward coordinate $x_1$ is drawn horizontally.
Both constructions yield resistance bounds of order
$\alpha^N/N^{D-1}$.}
\label{fig:resistance-flow}
\end{figure*}

Figure~\ref{fig:resistance-flow} illustrates the two network
constructions used to bound the true escape time.
For the lower bound, shorting each shell decreases effective
resistance and produces a series network with link
resistances $1/C_k$, as shown in
Figure~\ref{fig:resistance-flow}(a).

\begin{theorem}[Lower bound]
\label{thm:lower-bound}
Fix $D \ge 1$ and $0<p<1$, and let $\alpha=1/p>1$.
For the true Metropolis chain started from the origin,
\begin{equation}
\label{eq:lower-bound}
\E_0[\tau_N] \ge c_{D,p}\,\frac{\alpha^N}{N^{D-1}}
\qquad (N \ge 1),
\end{equation}
for some constant $c_{D,p} > 0$ independent of~$N$.
\end{theorem}

\begin{proof}
Shorting all vertices within each shell to a single node produces
the quotient network of Proposition~\ref{prop:quotient}.
Because $A_N$ separates the origin from the exterior of $B_N$,
the exterior does not affect the resistance to $A_N$: an equilibrium
potential is zero there. We may therefore perform this shorting in
$B_N$.
By Rayleigh monotonicity~\cite{doyle1984,levin2017}, shorting can only
decrease resistance:
$R_{\mathrm{eff}}(0, A_N) \ge R_{\mathrm{eff}}^{\mathrm{coll}}(0, N)$.
In the collapsed network, the conductance across cut
$k \leftrightarrow k+1$ is $C_k = p^{k+1}(2k+1)^{D-1}$
from~\eqref{eq:conductance}, so
\[
R_{\mathrm{eff}}^{\mathrm{coll}}(0,N)
= \sum_{k=0}^{N-1} \frac{1}{C_k}
= \sum_{k=0}^{N-1} \frac{\alpha^{k+1}}{(2k+1)^{D-1}}
= \Theta\!\left(\frac{\alpha^N}{N^{D-1}}\right).
\]
The lower bound in~\eqref{eq:hitting-bounds} then gives
$\E_0[\tau_N] \ge \mu(0)\,R_{\mathrm{eff}}(0,A_N)
\ge c_{D,p}\,\alpha^N/N^{D-1}$.
In fact, retaining only the final term of the resistance sum gives
\[
\E_0[\tau_N]\ge\frac{\alpha^N}{(2N-1)^{D-1}}
\ge 2^{-(D-1)}\frac{\alpha^N}{N^{D-1}},
\]
so one may take $c_{D,p}=2^{-(D-1)}$ for all $N\ge1$.
\end{proof}

\begin{remark}
Rayleigh monotonicity compares effective resistances. Combined with
Lemma~\ref{lem:hitting}, it gives the lower bound of order
$\alpha^N/N^{D-1}$ on $H_D(N)$ in Theorem~\ref{thm:lower-bound}.
\end{remark}

\subsection{Upper Bound via an Explicit Flow}
\label{sec:upper-bound}

We construct a unit current from the origin to a face of the target
shell. Thomson's principle then bounds the effective resistance by
the energy of this current. For a current $\theta$ on edges with
conductances $c(e)$, its energy is
\[
\mathcal E(\theta)=\sum_e \frac{\theta(e)^2}{c(e)},
\]
where each unoriented edge is counted once. The construction spreads
the current across progressively larger layers, reducing the energy
needed to reach the target.

\textbf{A large target face.}
Consider the face
\[
F_N^\circ
:=\{(N,y_2,\dots,y_D): |y_i|\le N-1,\ i=2,\dots,D\}
\]
of the target shell $A_N$. It contains
\[
|F_N^\circ|=(2N-1)^{D-1}=\Theta(N^{D-1})
\]
vertices. Since every trajectory that reaches this face has also
reached the shell,
\[
\E_0[\tau_N]\le \E_0[\tau_{F_N^\circ}].
\]
It therefore suffices to upper-bound the expected time to reach
$F_N^\circ$.

\textbf{A pyramid subnetwork.}
Write $n:=D-1$ and let
\[
Q_k:=\{-k,\dots,k\}^n,
\qquad |Q_k|=(2k+1)^n.
\]
We use the expanding layers
\[
L_k:=\{(k,y):y\in Q_k\},
\qquad k=0,\dots,N-1,
\]
followed by the target layer
\[
L_N^\circ:=\{(N,y):y\in Q_{N-1}\}=F_N^\circ.
\]
Within $L_k$, lateral edges have conductance $p^k/(2D)$.
The forward edges from $(k,y)$ to $(k+1,y)$ have conductance
$p^{k+1}/(2D)$.

Figure~\ref{fig:resistance-flow}(b) shows the construction
for $D=2$ and $N=4$. Current first advances along the
$x_1$ direction from $L_k$ to the embedded copy of $Q_k$
in $L_{k+1}$. It then spreads sideways to become uniform
on the larger cross-section $Q_{k+1}$.
The final forward step reaches $F_N^\circ$ without
further redistribution.
The following lemma bounds the energy required for
the sideways redistribution.

\begin{lemma}[Transport on nested cubes]
\label{lem:transport}
For integers $n\ge1$ and $k\ge0$, let $\rho_k$ be the uniform
probability distribution on $\{-k,\dots,k\}$, extended by zero
outside this interval. Let $u_k:=\rho_k^{\otimes n}$ be the
uniform probability distribution on $Q_k$. Here $u_k$ denotes a
probability distribution, distinct from the scalar proposal
probability in~\eqref{eq:uk}.

There is a nearest-neighbor current $\psi_k$ on $Q_{k+1}$ such that
\[
\operatorname{div}\psi_k=u_k-u_{k+1},
\]
where divergence means outgoing current minus incoming current
at each vertex. Thus the current removes the excess mass of $u_k$
and supplies the mass needed to obtain $u_{k+1}$. Moreover,
\begin{equation}
\label{eq:transport-energy}
\sum_{e\subset Q_{k+1}}\psi_k(e)^2
\le \frac{n}{(k+1)^n}.
\end{equation}
Each unoriented edge is counted once. The sum in
\eqref{eq:transport-energy} is the energy when every edge has
conductance one.
\end{lemma}

\begin{proof}
We first construct the current in one dimension, then apply this
construction one coordinate at a time.

In one dimension, orient every edge to the right. Define the
signed current from $j$ to $j+1$ by
\[
J_{j+1/2}
:=\sum_{t=-k-1}^{j}\bigl(\rho_k(t)-\rho_{k+1}(t)\bigr),
\qquad -k-1\le j\le k.
\]
A negative value means that the current travels to the left.
This definition has a simple interpretation: the current crossing
an edge equals the total excess mass to its left.

Set the current outside the interval to zero. Subtracting
successive cumulative sums gives
\[
J_{j+1/2}-J_{j-1/2}
=\rho_k(j)-\rho_{k+1}(j).
\]
Thus outgoing minus incoming current is exactly the required
change in mass at each vertex, including the endpoints.

Direct evaluation gives
\[
J_{j+1/2}
=-\frac{1}{2k+3}
+\frac{2(j+k+1)}{(2k+1)(2k+3)}.
\]
These currents increase from $-1/(2k+3)$ to $1/(2k+3)$, so
their absolute values are at most $1/(2k+3)$. There are $2k+2$
edges, and hence
\[
\sum_{j=-k-1}^{k}J_{j+1/2}^2
\le \frac{2k+2}{(2k+3)^2}
\le \frac{1}{k+1}.
\]

For general $n$, expand the distribution one coordinate at a time.
After expanding the first $r$ coordinates, the distribution is
\[
\nu^{(r)}
:=\rho_{k+1}^{\otimes r}\otimes
  \rho_k^{\otimes(n-r)},
\qquad r=0,\dots,n.
\]
In particular, $\nu^{(0)}=u_k$ and $\nu^{(n)}=u_{k+1}$.

At stage $r$, where $0\le r<n$, we expand coordinate $r+1$.
Fix the other coordinates, denoting their values by
$z=(z_i)_{i\ne r+1}$. The points obtained by varying coordinate
$r+1$ form a coordinate line. The total mass assigned to this
line is
\[
w_r(z)
:=\prod_{i=1}^{r}\rho_{k+1}(z_i)
  \prod_{i=r+2}^{n}\rho_k(z_i).
\]
On this line, use the one-dimensional currents multiplied by
$w_r(z)$: the current across its edge from $j$ to $j+1$ is
$w_r(z)J_{j+1/2}$.

Doing this on every such line defines a current $\psi^{(r)}$
using only edges in coordinate direction $r+1$. The
one-dimensional mass-balance identity shows that
\[
\operatorname{div}\psi^{(r)}
=\nu^{(r)}-\nu^{(r+1)}.
\]

To bound its energy, note that a uniform distribution on $m$
points has sum of squared probabilities equal to $1/m$.
Therefore
\[
\sum_z w_r(z)^2
=(2k+3)^{-r}(2k+1)^{-(n-r-1)}
\le (k+1)^{-(n-1)}.
\]
Here empty products are interpreted as $1$. Combining this with
the one-dimensional energy bound yields
\[
\begin{aligned}
\sum_e \psi^{(r)}(e)^2
&=\left(\sum_z w_r(z)^2\right)
  \left(\sum_{j=-k-1}^{k}J_{j+1/2}^2\right)\\
&\le (k+1)^{-(n-1)}(k+1)^{-1}\\
&=(k+1)^{-n}.
\end{aligned}
\]

Finally, add the currents from all $n$ stages:
\[
\psi_k:=\sum_{r=0}^{n-1}\psi^{(r)}.
\]
Their divergences telescope, giving
\[
\operatorname{div}\psi_k
=\sum_{r=0}^{n-1}
  \bigl(\nu^{(r)}-\nu^{(r+1)}\bigr)
=u_k-u_{k+1}.
\]
Each stage uses a different coordinate direction, so no edge is
used by two stages. Their energies therefore add:
\[
\sum_e\psi_k(e)^2
=\sum_{r=0}^{n-1}\sum_e\psi^{(r)}(e)^2
\le \frac{n}{(k+1)^n}.
\]
This also covers $k=0$.
\end{proof}

We now construct a unit flow whose energy gives the required
resistance upper bound. Recall that $\alpha=1/p>1$.

\begin{proposition}[Resistance upper bound]
\label{prop:resistance-upper}
Fix $D\ge1$ and $0<p<1$. There exists a finite constant
$C'_{D,p}$, independent of $N$, such that for every integer $N\ge1$,
\begin{equation}
\label{eq:Reff-upper}
R_{\mathrm{eff}}(0 \leftrightarrow F_N^\circ)
\le C'_{D,p}\,\frac{\alpha^N}{N^{D-1}}.
\end{equation}
\end{proposition}

\begin{proof}
For $D=1$, the target face is the single point $N$.
Send one unit of current along the path $0,1,\dots,N$.
The edge $k\leftrightarrow k+1$ has conductance $p^{k+1}/2$,
so this flow has energy
\[
2\sum_{k=0}^{N-1}\alpha^{k+1}
=\frac{2\alpha(\alpha^N-1)}{\alpha-1}
\le \frac{2\alpha}{\alpha-1}\alpha^N.
\]
Thomson's principle gives the claimed bound.

Now assume $D\ge2$ and write $n=D-1$.
We construct the flow through the pyramid layers described above.
At each layer $L_k$, the outgoing forward current will be
distributed uniformly over its $(2k+1)^n$ vertices.

\emph{Forward current.}
For $k=0,\dots,N-1$, put current $1/(2k+1)^n$ on each edge
from $(k,y)$ to $(k+1,y)$, with $y\in Q_k$.
These currents have total value one. Since each edge has
conductance $p^{k+1}/(2D)$, their total energy is
\[
E_k^{\mathrm{vert}}
=(2k+1)^n
 \frac{(2k+1)^{-2n}}{p^{k+1}/(2D)}
=\frac{2D\,\alpha^{k+1}}{(2k+1)^n}
\le \frac{2D\,\alpha^{k+1}}{(k+1)^n}.
\]

\emph{Redistribution within a layer.}
For $k=0,\dots,N-2$, the forward current arrives in $L_{k+1}$
uniformly on $Q_k$, whereas the next forward step requires it
to leave uniformly on $Q_{k+1}$.
Use the current $\psi_k$ from Lemma~\ref{lem:transport}
on the lateral edges of $L_{k+1}$ to make this redistribution.

The lemma gives
\[
\sum_e\psi_k(e)^2\le \frac{n}{(k+1)^n}.
\]
Each lateral edge in this layer has resistance
$2D\alpha^{k+1}$. Consequently, the redistribution energy satisfies
\[
E_{k+1}^{\mathrm{hor}}
=2D\alpha^{k+1}\sum_e\psi_k(e)^2
\le \frac{2Dn\,\alpha^{k+1}}{(k+1)^n}.
\]
The final forward step, from $L_{N-1}$ into $F_N^\circ$,
requires no further redistribution.

\emph{Conservation of current.}
At a vertex indexed by $y\in Q_j$ in an intermediate layer
$L_j$, where $1\le j\le N-1$, the incoming forward current
is $u_{j-1}(y)$ and the outgoing forward current is $u_j(y)$.
Here $u_{j-1}$ is extended by zero outside $Q_{j-1}$.
The lateral current has divergence
$u_{j-1}(y)-u_j(y)$, so the total divergence is
\[
-u_{j-1}(y)+u_j(y)
+\bigl(u_{j-1}(y)-u_j(y)\bigr)=0.
\]
The origin supplies one unit of current, and the target face
receives one unit, uniformly distributed over its vertices.
Thus the construction is a unit flow from $0$ to $F_N^\circ$.
For $N=1$, it consists of just the edge from the origin to
the target point. Set the current to zero on every edge
outside the construction.

\emph{Total energy.}
Forward edges and lateral edges are disjoint, and different
redistribution steps use different layers. Their energies
therefore add. By Thomson's principle,
\[
\begin{aligned}
R_{\mathrm{eff}}(0 \leftrightarrow F_N^\circ)
&\le \sum_{k=0}^{N-1}E_k^{\mathrm{vert}}
   +\sum_{k=0}^{N-2}E_{k+1}^{\mathrm{hor}}\\
&\le 2D(1+n)\sum_{j=1}^{N}\frac{\alpha^j}{j^n}.
\end{aligned}
\]

It remains to bound this sum by a constant times its final term.
Set $a_j:=\alpha^j/j^n$. Since
\[
\frac{a_{j-1}}{a_j}
=\alpha^{-1}\left(\frac{j}{j-1}\right)^n
\longrightarrow \alpha^{-1}<1,
\]
there exist $q\in(0,1)$ and an integer $J\ge1$,
depending only on $n$ and $p$, such that
$a_{j-1}\le q a_j$ for every $j\ge J+1$.
For $N\ge J$, repeated application of this inequality gives
\[
\sum_{j=J}^{N}a_j
\le a_N\sum_{\ell=0}^{N-J}q^\ell
\le \frac{a_N}{1-q}.
\]
Also $a_N\ge a_J$, so
\[
\sum_{j=1}^{J-1}a_j
\le \frac{\sum_{j=1}^{J-1}a_j}{a_J}\,a_N.
\]
Combining these bounds, and enlarging the constant to cover
the finitely many cases $N<J$, yields
\[
\sum_{j=1}^{N}\frac{\alpha^j}{j^n}
\le K_{n,p}\frac{\alpha^N}{N^n}.
\]
Since $n=D-1$, this proves the proposition.
\end{proof}

We can now state and prove the full scaling law.

\begin{theorem}[True-chain escape-time scaling]
\label{thm:true-scaling}
Fix $D \ge 1$ and $0<p<1$, and let $\alpha = 1/p$.
For the true Metropolis chain on $\Z^D$ with energy
$V(x) = \norm{x}_\infty$, started from the origin,
\begin{equation}
\label{eq:true-scaling}
\E_0[\tau_N] = \Theta\!\left(\frac{\alpha^N}{N^{D-1}}\right)
\qquad (N \to \infty).
\end{equation}
\end{theorem}

\begin{proof}
The lower bound is Theorem~\ref{thm:lower-bound}.
For the upper bound, since $\tau_N \le \tau_{F_N^\circ}$,
Lemma~\ref{lem:hitting} gives
{\small
\[
\E_0[\tau_N] \le \E_0[\tau_{F_N^\circ}]
\le M_{D,p}\,R_{\mathrm{eff}}(0 \leftrightarrow F_N^\circ)
\le M_{D,p}\,C'_{D,p}\,\frac{\alpha^N}{N^{D-1}},
\]
\par}
using Proposition~\ref{prop:resistance-upper}.
Since $M_{D,p} < \infty$ depends only on $D$ and $p$, combining
with the lower bound yields~\eqref{eq:true-scaling}.
\end{proof}

\begin{remark}
The shell-averaged chain identifies the correct bottleneck scale:
the resistance across shell $k$ is of order $\alpha^k/k^{D-1}$.
The lower bound shows this is unavoidable (collapsing shells only
helps), while the explicit face flow shows that the true chain can
exploit the full $(D-1)$-dimensional surface area of each shell,
so no larger asymptotic obstruction is present.
\end{remark}


\section{Numerical Results}

\subsection{Numerical Method}
We compare the true Metropolis mean escape time $H_D(N)$
with the shell-quotient prediction $Q_D(N)$ for $D=1,2,3$.
Both count attempted moves, including rejected proposals.
We compute $H_D(N)$ by grouping states that differ only in
coordinate signs or order and solving the resulting sparse
linear systems for escape probabilities and excursion times.
This symmetry reduction preserves the true dynamics exactly.
We evaluate $Q_D(N)$ using the closed-form expression in Theorem~\ref{thm:exact-escape}.

\subsection{Dimensional Speedup}
\begin{figure*}[!b]
\centering
\includegraphics[width=0.8\textwidth]{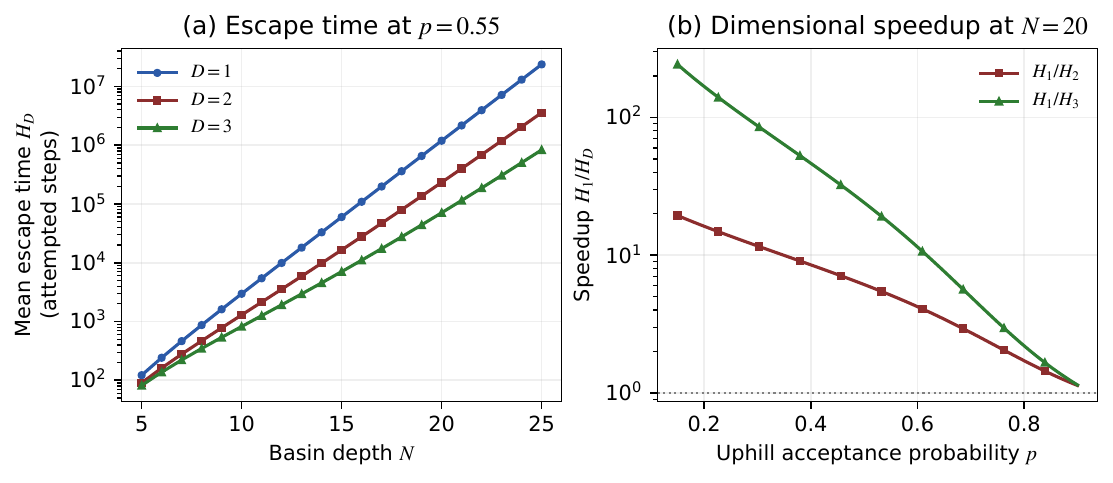}
\caption{Dimensional dependence of the true Metropolis escape time.
(a) Mean escape time versus basin depth at $p=0.55$, for
$D=1,2,3$ and integer depths $N=5,\ldots,25$.
(b) Speedup relative to one dimension, $H_1/H_D$, at $N=20$
for 50 acceptance probabilities between $0.15$ and $0.9$.
Values above one indicate faster escape in dimension $D$.
Both vertical axes are logarithmic.}
\label{fig:vi-dimensional-speedup}
\end{figure*}
Figure~\ref{fig:vi-dimensional-speedup}(a) shows that higher
dimension reduces escape time in the basins tested at $p=0.55$.
The separation grows with depth: at $N=25$, escape is about
$6.6$ times faster in two dimensions and $28.4$ times faster
in three dimensions than in one dimension.
This illustrates the polynomial dimensional advantage in the
proved scaling $H_D(N)=\Theta(p^{-N}/N^{D-1})$.

Figure~\ref{fig:vi-dimensional-speedup}(b) shows how the
speedup depends on the acceptance probability at $N=20$.
The three-dimensional walk escapes about $243$ times faster
than the one-dimensional walk at $p=0.15$, but only about
$1.14$ times faster at $p=0.9$.
The dimensional advantage is therefore much larger at strong
uphill rejection in this example.
For shallow basins, the ordering can reverse: at $N=2$ and
$p=0.9$, $H_1\approx4.57$ whereas $H_2\approx5.02$.

\subsection{Accuracy of the Shell Quotient}
\begin{figure*}[!t]
\centering
\includegraphics[width=\textwidth]{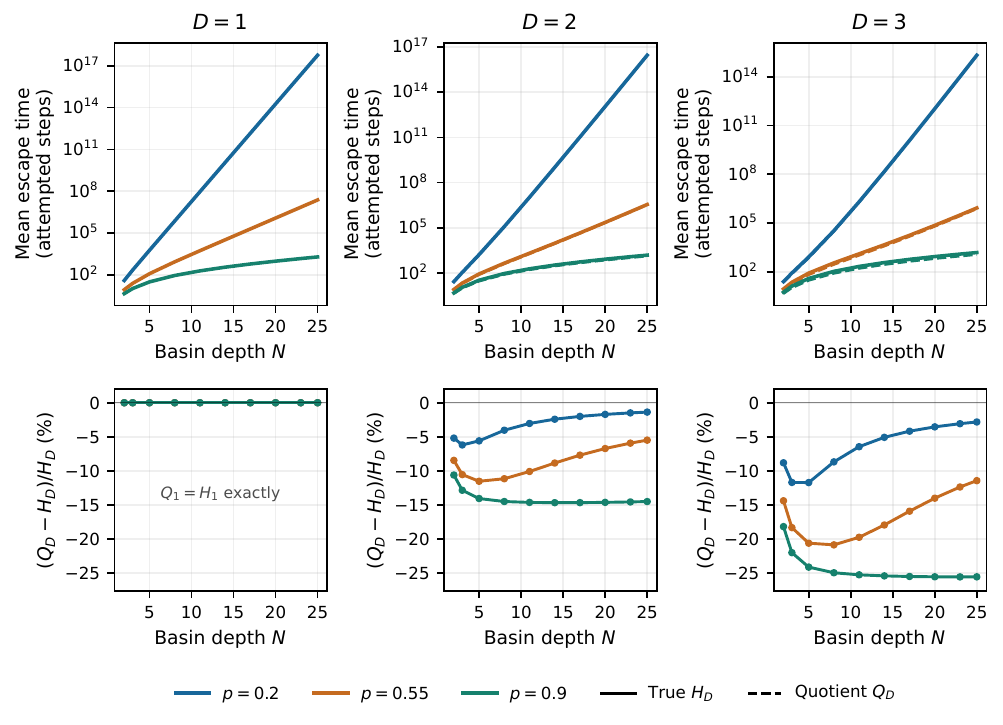}
\caption{True Metropolis mean escape time $H_D(N)$ (solid)
and shell-quotient mean $Q_D(N)$ (dashed) versus basin depth
(top), with signed relative discrepancy
$100(Q_D/H_D-1)$ in percent (bottom).
Columns correspond to $D=1,2,3$; colors indicate
$p=0.2,0.55,0.9$.
The computed depths are $N=2,3,5,8,11,14,17,20,23,25$.
Times count attempted moves, and the upper panels use
logarithmic vertical axes. For $D=1$, the means coincide
exactly and only the solid curves are shown.}
\label{fig:vi-quotient-comparison}
\end{figure*}
The upper panels of Figure~\ref{fig:vi-quotient-comparison}
compare $Q_D(N)$ with $H_D(N)$.
Their curves largely overlap on the logarithmic scale, so
the lower panels show the signed relative discrepancy
\[
\delta_D(N,p):=100\left(\frac{Q_D(N)}{H_D(N)}-1\right).
\]
A negative value means that the quotient underestimates the
true mean escape time. In one dimension, $Q_1=H_1$ exactly.

At $N=25$ and $p=0.2$, the quotient underestimates the true
escape time by approximately $1.4\%$ in two dimensions and
$2.8\%$ in three dimensions.
At $p=0.9$, the corresponding discrepancies are larger:
$14.5\%$ and $25.5\%$.
At low $p$, the discrepancy decreases over the larger depths
shown, whereas at $p=0.9$ it changes little at those depths.
The quotient captures the escape-time scale, while its
quantitative accuracy depends on dimension, acceptance
probability, and basin depth.

The discrepancy persists even without uphill rejection:
at $p=1$ and $N=40$, numerical calculations give
$H_2/N^2\approx1.1786$ and $H_3/N^2\approx1.3488$,
corresponding to quotient discrepancies of approximately
$-15.15\%$ and $-25.86\%$, respectively.
These values provide context for the nearly flat discrepancy
curves at $p=0.9$, without establishing a common asymptotic limit.

\section{Conclusion}

In this paper, we have extended the finite-time analysis of
constant-temperature simulated annealing from one dimension to
$D$-dimensional cubic basins defined by the infinity norm.
We derived a closed-form expression for the mean escape time of the
shell-averaged Metropolis chain (Theorem~\ref{thm:exact-escape}),
proved that its asymptotic scaling is $\Theta(\alpha^N/N^{D-1})$
(Theorem~\ref{thm:asymptotic}), and established that the true
Metropolis chain, without shell averaging, has the same scaling
(Theorem~\ref{thm:true-scaling}), via a lower bound from Rayleigh
monotonicity and a matching upper bound from an explicit Thomson flow.

The central conclusion is specific to deep cubic basins: for fixed
$D$ and $0<p<1$, the mean time to first reach the outer shell
(any site with $\norm{x}_\infty=N$) is
$\Theta(p^{-N}/N^{D-1})$ attempted steps. The polynomial factor reflects
the available boundary area and yields an asymptotic dimensional speedup
within this family. 

Several directions for future work remain.
First, extending the analysis to non-cubic basins (e.g., basins
defined by the $\ell^1$ or $\ell^2$ norm) would test the robustness
of the polynomial speedup to basin geometry.
Second, the multi-basin analysis of~\cite{ramachandran2026} could be
extended to investigate escape and global hitting times in
multidimensional multi-basin landscapes.
 Third, studying continuous-state counterparts could test whether the
discretization effects reported in~\cite{ramachandran2026} persist
in higher dimensions. Fourth, determining a leading constant for the true-chain escape time would sharpen Theorem~\ref{thm:true-scaling}; the constant in Remark~\ref{rmk:sharp} currently applies only to the quotient chain.

\section*{Code and expanded proofs.}
The numerical notebook and expanded proofs in Lamport's
hierarchical proof format are available at
\url{https://github.com/ANRGUSC/metropolis-highD/}.
The notebook reproduces the numerical results and figures,
while the supplement makes intermediate claims and dependencies
explicit to assist with proof verification.
These structured proofs are not machine-checked formal proofs.

\section*{AI Use Statement}
The authors made substantial use of ChatGPT, Claude, and Grok
in developing the mathematical results, numerical computations,
illustrations, exposition, and formatting of this paper.
The authors accept full responsibility for the paper's contents.


\end{document}